\documentclass[10pt, a4paper, twocolumn, twoside]{IEEEtran}

\usepackage[utf8]{inputenc}
\usepackage[T1]{fontenc}
\usepackage{url}
\usepackage{ifthen}
\usepackage{cite}
\usepackage{cancel}

\usepackage{pdfsync}

\usepackage[top=20mm, left=1.45cm, right=1.45cm, bottom=3.0cm]{geometry}
\usepackage{cite}
\usepackage{enumerate}
\usepackage{graphicx}
\usepackage[cmex10]{amsmath} % prevents amsmath from using a Type 3 font for math within footnotes
\usepackage{amssymb}
\usepackage{mathtools}
\usepackage{xspace}
\usepackage{subfigure}
\usepackage[lined,linesnumbered,ruled,commentsnumbered]{algorithm2e} % for algorithm
\usepackage{colonequals}
\usepackage[mathscr]{euscript}
\usepackage{fixltx2e}    % for text subscript
\usepackage{amsthm}
\usepackage{mathrsfs}
\usepackage[belowskip=-25pt,aboveskip=5pt]{caption}

\usepackage{float}

\usepackage{epsfig}
\usepackage{epstopdf}

\usepackage{tikz}
\usepackage{pgfplots}
\pgfplotsset{colormap/jet}

\usetikzlibrary{arrows,shapes,backgrounds,plotmarks,positioning}

\pgfplotsset{compat=newest}                         % move axis labels close to the tick label automatically
\pgfplotsset{plot coordinates/math parser=false}
\newlength\figureheight
\newlength\figurewidth

\DeclareMathOperator*{\argmax}{arg\,max}

\newtheorem{theorem}{Theorem}%[section]

\newtheorem{corollary}{Corollary}

\newtheorem{example}{Example}

\newcommand{\E}{\mathbb{E}}

\newcommand{\Set}[1]{\{#1\}}

\newcommand{\X}{\mathcal{X}}

\newcommand{\Y}{\mathcal{Y}}
\renewcommand{\S}{\mathcal{S}}
\newcommand{\eps}{\epsilon}
\newcommand{\Xepsc}{\X_\epsilon^c}
\newcommand{\Xeps}{\X_\epsilon}
\newcommand{\Xepsalpc}{\X_{\eps}^c}
\newcommand{\Xepsalp}{\X_{\eps}}
\newcommand{\Xepsbar}{\X_{\bar{\epsilon}}}
\newcommand{\Xepscbar}{\X_{\bar{\epsilon}}^c}

\newcommand{\Xdepscbar}{\X_{\eps,\bar{\eps}}^c}

\begin{document}
\pagestyle{empty}
\thispagestyle{empty}

\title{$\alpha$-Information-theoretic Privacy Watchdog and Optimal Privatization Scheme}
\thispagestyle{empty}
%% Many authors with many affiliations:
 \author{Ni Ding, Mohammad Amin Zarrabian and Parastoo Sadeghi\thanks{N.~Ding is with the School of Computing and Information Systems, University of Melbourne, Melbourne. M.~A.~Zarrabian is with the ANU College of Engineering and Computer Science at the Australian National University. P.~Sadeghi is with the School of Engineering and Information Technology, University of New South Wales, Canberra. The work of P.~Sadeghi  and M.~A.~Zarrabian is supported by the ARC Future Fellowship FT190100429 and partly by the Data61 CRP: IT-PPUB. Emails: \texttt{ni.ding@unimelb.edu.au, u7139330@anu.edu.au, %m.amin.zarrabian@gmail.com
 p.sadeghi@unsw.edu.au}.}}

\maketitle

\begin{abstract}
This paper proposes an $\alpha$-lift measure for data privacy and determines the optimal privatization scheme that minimizes the $\alpha$-lift in the watchdog method. To release data $X$ that is correlated with sensitive information $S$, the ratio $l(s,x) = \frac{p(s|x)}{p(s)} $ denotes the `lift' of the posterior belief on $S$ and quantifies data privacy.  The $\alpha$-lift is proposed as the $L_\alpha$-norm of the lift: $\ell_{\alpha}(x) = \| l(\cdot,x) \|_{\alpha} = (\E[l(S,x)^\alpha])^{1/\alpha}$. This is a tunable measure: When $\alpha < \infty$, each lift is weighted by its likelihood of appearing in the dataset (w.r.t. the marginal probability $p(s)$); For $\alpha = \infty$, $\alpha$-lift reduces to the existing maximum lift.
To generate the sanitized data $Y$, we adopt the privacy watchdog method using $\alpha$-lift: Obtain $\Xepsalpc$ containing all $x$'s such that $\ell_{\alpha}(x) > e^{\eps}$; Apply the randomization $r(y|x)$ to all $x \in \Xepsalpc$, while all other $x \in \X \setminus \Xepsalpc$ are published directly.
For the resulting $\alpha$-lift $\ell_{\alpha}(y)$, it is shown that the Sibson mutual information $I_{\alpha}^{S}(S;Y)$ is proportional to $\E[ \ell_{\alpha}(y)]$. We further define a stronger measure $\bar{I}_{\alpha}^{S}(S;Y)$ using the worst-case $\alpha$-lift: $\max_{y} \ell_{\alpha}(y)$.
We prove that the optimal randomization $r^*(y|x)$ that minimizes both $I_{\alpha}^{S}(S;Y)$ and $\bar{I}_{\alpha}^{S}(S;Y)$ is $X$-invariant, i.e., $r^*(y|x) = R(y), \forall x\in \Xepsalpc$ for any probability distribution $R$ over $y \in \Xepsalpc$.
%
% Numerical experiments show that $\alpha$-lift can provide a more flexible privacy-utility tradeoff than the existing maximum lift.
Numerical experiments show that $\alpha$-lift can provide flexibility in the privacy-utility tradeoff.
\end{abstract}

\section{Introduction}

The world has never been more interconnected. Being increasingly reliant on the internet, we are also exposed to more cyber-threats and cyber-crimes that try to infiltrate networks, hijack systems and steal identities and confidential data.
The scale and sophistication of cyberthreats continue to increase, highlighting the importance of ensuring a safe and secure online data sharing environment.
For any malicious data analytics, the data disclosure risk can be quantified by statistical measures. For example, the differential privacy in \cite{Dwork2014DP} and local differential privacy in \cite{LDP2013MiniMax} quantify the relative change in the output probability (e.g., of a randomization scheme) conditioned on two distinct values of the sensitive data. The smaller the upper bound on this change, the more private the released data.

In information theory, data privacy is measured by the difficulty for an adversary to infer/estimate the sensitive information \cite{PvsInfer2012}. Denote by $X$ the data to be released to the public. Let $S$ be the private attribute of $X$. We want to randomize/encode $X$ to $Y$ such that publishing $Y$ protects the privacy of $S$.
That is, the relative change of the posterior belief $p(s|y)$ from the prior $p(s)$, denoting the adversary's knowledge gain on the sensitive data $S$, needs to be restricted.
The ratio $l(s,y) = \frac{p(s|y)}{p(s)}$, called \emph{lift} in \cite{Watchdog2019}, appears in many information-theoretic data privacy measures, e.g., the average, maximal and $\alpha$-leakage in \cite{PvsInfer2012}, \cite{Issa2016MaxL} and \cite{Liao2019Alpha}, respectively.\footnote{The logarithm of the lift, the log-lift or more widely known as information density, is used in \cite{PvsInfer2012,Issa2016MaxL,Liu2020ISIT,Liao2019Alpha}. The average leakage in \cite{PvsInfer2012} is the expectation of the log-lift $\E [\log l(S,Y)]$, which is in fact the mutual information $I(S;Y)$. The maximal leakage in \cite{Issa2016MaxL,Liu2020ISIT} and the $\alpha$-leakage in \cite{Liao2019Alpha} use variations of the log-lift $\log\frac{\E[\max_{s} p(s|Y)]}{\max_{s} p(s)}$ and $\log \frac{\E[\| p_{S|Y}(\cdot|Y) \|_{\alpha} ]}{\| p_S(\cdot)\|_{\alpha}}$, respectively.}
The maximum lift $\max_{s,y} l(s,y)$ is studied in \cite{Watchdog2019} as a stronger notion of privacy in that it upper bounds the mutual information, Sibson and Arimoto mutual information  and the local differential privacy \cite[Proposition~1]{Watchdog2019}.

To privatize $X$, a watchdog randomization scheme is proposed in \cite[Section~III]{Watchdog2019}. The idea is to censor the lift $l(s,x)$ for each element $x \in \X$ in the original dataset. A positive threshold $\eps$ is applied to filer out a set $\Xepsalpc$ containing all `high-risk' elements $x$'s.\footnote{In \cite{Watchdog2019,Sadeghi2020ITW}, `high risk' elements refers to $x$ such that the absolute value of log-lift exceeds $\eps$, i.e., $\Xepsalpc = \Set{ x \in \X \colon \max_{s \in \S}|\log l(s,x)| > \eps} $.}
The randomization $r(y|x)$ only applies to $x \in \Xepsalpc$, while all `low risk' elements $x \in \X \setminus \Xepsalpc$ are directly released without any distortion.
The optimal randomization $r^*(y|x)$ that minimizes the maximum lift for all $y\in\Xepsc$ was proved to be \emph{$X$-invariant} in a recent study~\cite{Sadeghi2020ITW}. This study also highlights the \emph{two-fold interpretation} of the joint probability $p(s,x)$ and the marginals $p(s)$ and $p(x)$: They not only determine the value of lift, but also indicate the chances this lift value appears in the dataset.
The fact is that a high risk event may occur with least probably, e.g., for some $x$, the worst-case lift $\max_{s \in \S} l(s,x)$ could happen with only probability $p(S \in \argmax_{s \in \S} l(s,x)) = 10^{-3}$.
Therefore, it is desirable to have a lift measure taking into consideration the probability distribution in the original dataset.

In this paper, we propose an $\alpha$-lift measure as the $L_\alpha$-norm over the probability space with the measure being the marginal probability $p(s)$. For each $x \in \X$, regarding the lift $l(s,x)$ for all $s \in \S$ as an $|\S|$-dimensional vector, the $\alpha$-lift is defined as $\ell_{\alpha}(x) = \| l(\cdot,x) \|_{\alpha} = (\E[l(S,x)^\alpha])^{1/\alpha}$.
This measure is tunable by $\alpha$: For $\alpha \in (1, \infty)$, the $\alpha$-exponential of each lift is weighted by $p(s)$ denoting the likelihood it appears in the dataset; When $\alpha = \infty$, $\ell_{\infty}(x) = \max_{s} l(s,x)$ denotes the maximum lift that is considered in \cite{Watchdog2019,Sadeghi2020ITW}.
We show that the Sibson mutual information is a function of the expectation of $\alpha$-lift: $I_{\alpha}^{S} (S;X) = \frac{\alpha}{\alpha-1} \log \E[\ell_{\alpha}(X)]$ for all $\alpha \in (1,\infty)$.
Replacing $\E[\cdot]$ by $\max_{x \in \X}$, we propose the maximum Sibson mutual information $\bar{I}_{\alpha}^{S} (S;X)$ to quantify the worst-case data privacy risk over $\X$.
We adopt the privacy watchdog method using the $\alpha$-lift for the design of the privatization scheme. That is, the `high risk' set $\Xepsalpc$ is comprised of all instances $x$ such that $\log \ell_{\alpha}(x) > \eps$.
The optimality of a privatization scheme $r(y|x)$ is determined by the resulting $\alpha$-lift $\ell_{\alpha}(y)$ for all $y \in \Xepsalpc$. We prove that the randomization $r^*(y|x)$ that minimizes both $\max_{y \in \Xepsalpc} \ell_{\alpha}(y)$ and $\sum_{y \in \Xepsalpc} p(y)\ell_{\alpha}(y)$ is $X$-invariant. That is, $r^*(y|x) = R(y), \forall x \in \Xepsalpc$, where $R(y)$ is any distribution over $y \in \Xepsalpc$ such that $\sum_{y \in \Xepsc} R(y) = 1$.
This, in turn, proves that an $X$-invariant $r^*(y|x)$ also minimizes $I_{\alpha}^S(S;Y)$ and $\bar{I}_{\alpha}^S(S;Y)$.

\section{$\alpha$-lift}

Let $S$ and $X$ be the random variables having countable alphabets $\S$ and $\X$, respectively. We describe the privacy measures resulted from the joint probability $p(s,x)$.
For each $x \in \X$, $p(s|x)$ denotes the posterior belief on the sensitive data $S$, and the ratio
\begin{equation} \label{eq:Lift}
    l(s,x) \triangleq \frac{p(s|x)}{p(s)} = \frac{p(s,x)}{p(s)p(x)}
\end{equation}
for all $s \in \S$ measures the relative change of this posterior belief from the prior for each instance $x$. This measure is called the \emph{lift} in \cite{Watchdog2019} for that it indicates an adversary's knowledge gain on the private data when $x$ is present in the released data.
%
%
%
%The logarithm of the lift appears in many information-theoretic privacy measures, e.g., the mutual information, interpreted as the average privacy leakage, in \cite{PvsInfer2012}; maximal leakage \cite{MaxL2020} and $\alpha$-leakage \cite{Liao2019Alpha}.
%%
%It is shown in \cite[Lemma~1, Propostion~1]{Watchdog2019} \cite{PvsInfer2012} that if $\max_{(s,x) \in \S \times \X} |\log l(s,x)| = \epsilon$, then local differential privacy and Arimoto and Sibson mutual information are upper bounded.

\emph{Watchdog method}: In~\cite{Sadeghi2020ITW,Watchdog2019}, a lift-based randomization was proposed based on the log-lift or the \emph{information density} $i(s,x) = \log l(s,x)$.\footnote{We use natural logarithm in this paper.}
The alphabet $\X$ is bi-partitioned into $\Set{\Xeps, \Xepsc}$, where $\Xepsc = \Set{ x \in \X \colon \max_{s \in \S} |i(s,x)| > \epsilon}$ and $\Xeps = \X \setminus \Xepsc$. The set $\Xepsc$ is randomized, while $\Xeps$ is published without any perturbation.
In this method, $\Xeps$ is determined by the maximum lift over all $s \in \S$ for each $x \in \X$. Below, we use the concept of $L_\alpha$-norm on the probability space to propose a relaxed lift measure on $\X$.

%But, a cogent argument is that this worst value $\max_{s} |i(s,x)|$ may not appear very often in the dataset.
%%
%As pointed out in \cite[Section V-A]{Sadeghi2020ITW}, a pair $(s,x)$ could have large $l(s,x)$ but very small $p(s,x)$.\footnote{This is the case when the marginals $p(s)$ and $p(x)$ are both small. Also see Example~\ref{ex:Motivation}. In this case, it is more urgent to privatize the pairs $(s,x)$ with high $p(s,x)$, but not necessarily incur the worst-case log-lift. }
%In this case, it is desirable if the privacy measure can take into account the probability, e.g., $p(s,x)$, $p(s)$ or $p(x)$ denoting the chances the lift value $l(s,x)$ appears in the original dataset.

For $\alpha \in (1, \infty)$, define the \emph{$\alpha$-lift}
\begin{equation} \label{eq:AlphaLiftX}
	\ell_{\alpha}(x) \triangleq  \left(\sum_{s \in \S}p(s)\left(\frac{p(s, x)}{p(s) p(x)}\right)^{\alpha}\right)^{1 /\alpha}, \quad \forall x \in \X.
\end{equation}
When $\alpha < \infty$, for each $x \in \X$, the value $\ell_{\alpha}(x)$ depends on the lift $l(s,x)$, as well as the marginal probability $p(s)$ for all $s \in \S$.
As $\alpha \rightarrow \infty$, $\ell_{\infty} (x)  = \max_{s \in \S} l(s,x)$ reduces to the maximum lift that is used in \cite{Sadeghi2020ITW,Watchdog2019}.
Note that, when $\alpha < \infty$, an element $x^*$ that incurs the largest lift $x^* \in \argmax_{x \in \X} \ell_{\infty} (x)$ does not necessarily have the maximum $\alpha$-lift.

\begin{example} \label{ex:Motivation}
Consider the conditional probability $p(x|s)$ in Table~\ref{tab:Example}. For the marginal probability $p(S = 1) = 0.6$ and $p(S = 2) = 0.4$, the maximum lift occurs at $X = d$, i.e., $\max_{x \in \Set{a,\dotsc,d}} \ell_{\infty}(x) = \ell_{\infty}(X = d) = 1.8182$.
    \begin{table}[H]
        \vspace{20pt}
        \caption{An Example of $p(x|s)$} \label{tab:Example}
        \begin{center}
        \begin{tabular}{ccccc}
            \hline\hline
            & $X = a$ & $X=b$ & $X=c$ & $X = d$ \\ \hline
            $S = 1$ & 0.2  & 0.05 & 0.7 & $0.05$  \\ \hline
            $S = 2$ & 0.6  & 0.1 & 0.1 & $0.2$\\  \hline
            \end{tabular}
        \end{center}
    \end{table}
But, when $\alpha = 1.5$, the maximum $\alpha$-lift is $\max_{x \in \Set{a,\dotsc,d}} \ell_{1.5}(x) = \ell_{1.5}(X = c) = 1.1084$.
See also Appendix~\ref{app:example} for more details on how $\alpha$-lift varies with $p(s)$.
\end{example}

\emph{Sibson mutual information in terms of $\alpha$-lift}: The $\alpha$-lift well interprets the Sibson mutual information \cite{Verdu2015AlphaMI} in data privacy
\begin{equation}
\begin{aligned}
		I_{\alpha}^{S}(S;X) &\triangleq  \frac{\alpha}{\alpha-1} \log \sum_{x} \left(  \sum_{s} p(s) p (x|s)^{\alpha} \right)^{1/\alpha} \label{eq:Sibson} \\
		& = \frac{\alpha}{\alpha-1} \log \sum_{x} p(x) \left(  \sum_{s} p(s) \Big( \frac{p(s,x)}{p(s)p(x)} \Big)^{\alpha} \right)^{1/\alpha}  \\
        & = \frac{\alpha}{\alpha-1} \log \E[\ell_{\alpha}(X)] , \quad \forall \alpha \in (1,\infty).
\end{aligned}
\end{equation}
$I_{\alpha}^{S}(S;X)$ measures the expected $\alpha$-lift w.r.t. the marginal distribution of $X$. This also relates  the lift and log-lift in~\cite{Sadeghi2020ITW,Watchdog2019} to the Sibson mutual information for $\alpha = \infty$: $I_{\infty}^{S}(S;X) = \log \E[\ell_{\infty}(X)] = \log \E[ \max_{s \in \S} l(s,X)] = \log \E[\max_{s \in \S} \exp(i(s,X)) ]$ (see \cite[eq~(56)]{Verdu2015AlphaMI}).
The worst-case data privacy over $\X$ can be measured by replacing the expectation by the maximization in the Sibson mutual information
\begin{align}
		\bar{I}_{\alpha}^{S}(S;X) \triangleq  \frac{\alpha}{\alpha-1} \log \max_{x \in \X} \ell_{\alpha}(x) , \quad \forall \alpha \in (1,\infty).
\end{align}
We call $\bar{I}_{\alpha}^{S}(S;X)$ the maximum Sibson mutual information in this paper.
Then, the maximum log-lift can be denoted by $\bar{I}_{\infty}^{S}(S;X) = \log \max_{x \in \X} \ell_{\infty}(X) = \max_{(s,x) \in \S \times \X} i(s,x)$.
It is clear that if $\bar{I}_{\infty}^{S}(S;X) = \eps$,
\[ I_{\alpha}^{S}(S;X) \leq \bar{I}_{\alpha}^{S}(S;X) \leq  \frac{\alpha}{\alpha-1} \eps, \quad \forall \alpha \in (1, \infty). \]
This generalizes the lift-based upper bound on $I_{\alpha}^{S}(S;X)$ in \cite[Propostion~1]{Watchdog2019}.\footnote{It is shown in \cite[proof of Propostion~1]{Watchdog2019} that if $i(s,x) \leq \eps, \forall x \in \X$, $I_{\alpha}^{S}(S;X) \leq \frac{\alpha}{\alpha-1} \eps$.}

\section{Optimal Randomization}

Denote a data randomization scheme by the transition probability $p(y|x), \forall x, y$. This is a Markov randomization method: $p(y|s,x) = p(y|x),  \forall s, x, y$. It forms the Markov chain $S - X - Y$, which is the basic model for data privacy studies in \cite{Ding2019ITW,PF2014,Issa2016MaxL,Liao2019Alpha,Liu2020ISIT}.
Each randomization results in a posterior belief $p(s|y)$, by which $l(s,y) = \frac{p(s|y)}{p(s)} = \frac{p(y|s)}{p(y)}$, where $p(y|s) = \sum_{x} p(y|x) p(x|s)$ and $p(y) = \sum_{x} p(y|x) p(x)$. The resulting $\alpha$-lift of $y\in \Y$ is

%\begin{equation*}
%	   \ell_{\alpha}(y) \triangleq  \left(\sum_{s \in \S}p(s)\left(\frac{p(y|s)}{p(y)}\right)^{\alpha}\right)^{1 /\alpha}  \forall \alpha \in (1, \infty).
%\end{equation*}
\begin{equation*}
    \begin{aligned}
	   \ell_{\alpha}(y) &\triangleq  \left(\sum_{s \in \S}p(s)\left(\frac{p(s, y)}{p(s) p(y)}\right)^{\alpha}\right)^{1 /\alpha} \\
                        &= \left(\sum_{s \in \S}p(s)\left(\frac{\sum_{x}p(y|x)p(x|s)}{ \sum_{x} p(y|x)p(x)}\right)^{\alpha}\right)^{1 /\alpha}, \ \forall \alpha \in (1, \infty).
    \end{aligned}
\end{equation*}

\subsection{$\alpha$-privacy Watchdog Method}

Let $i_{\alpha}(x) = \log \ell_{\alpha}(x)$. For two instances $x,x' \in \X$ such that $i_{\alpha}(x) \leq i_{\alpha}(x')$, releasing $x$ is more privacy-preserving than $x'$.
We say that \emph{$(\eps,\alpha)$-log-lift} is attained at $x$ if $i_{\alpha}(x) \leq \eps$.
We adopt a watchdog randomization scheme in \cite{Sadeghi2020ITW,Watchdog2019} through using the logarithm of $\alpha$-lift as follows.
For a threshold $\eps > 0$, denote
\begin{equation}
    \Xepsalpc = \Set{ x \in \X \colon i_{\alpha}(x) > \eps }.
\end{equation}
Let $\Xepsalp = \X \setminus \Xepsalpc$.
The threshold $\eps$ determines the cut $\Set{\Xepsalp, \Xepsalpc}$, which categorizes the alphabet $\X$ into two subsets: the $x$'s that attain $(\eps,\alpha)$-log-lift and those that do not.
To design the randomization, we only need to treat those `high risk' $x$'s in $\Xepsalpc$, while the `low risk' ones $\Xepsalp$ can be directly released.
That is, we consider the randomization
\begin{equation} \label{eq:Mechanism}
    p(y|x) = \begin{cases} 1 & x = y \in \Xepsalp, \\
                           0 & x,y \in \Xepsalp, x \neq y,\\
                           r(y|x) & x, y \in \Xepsalpc, \end{cases}
\end{equation}
where for each $x \in \Xepsalpc$, $\sum_{y \in \X} r(y|x) = \sum_{y \in \Xepsalpc} r(y|x) = 1$.
We call this randomization the \emph{$\alpha$-privacy watchdog method}, which results in the $\alpha$-lift
\begin{multline*}
    \ell_{\alpha}(y) = \\
    \begin{cases} \left(\sum_{s \in \S}p(s)\left(\frac{p(s,x)}{p(s) p(x)}\right)^{\alpha}\right)^{1 /\alpha} &\hspace{-4mm} x,y \in \Xepsalp \colon x = y,\\
                                     \left(\sum_{s \in \S}p(s)\left(\frac{\sum_{x \in \Xepsalpc }r(y|x)p(x|s)}{ \sum_{x \in \Xepsalpc} r(y|x)p(x)}\right)^{\alpha}\right)^{1 /\alpha} & y \in \Xepsalpc.
                                     \end{cases}
\end{multline*}

\subsection{Optimality of $X$-invariant Randomization}
A good privacy-preserving randomization should generate a random variable $Y$ that reduces the $\alpha$-lift. For the $\alpha$-privacy watchdog method in \eqref{eq:Mechanism}, it suffices to search an $r(y|x)$ that minimizes the $\alpha$-lift in $\Xepsalpc$.
We show below that the optimal scheme is $X$-invariant.

Consider a special type of randomization that for each $y \in \Xepsalpc$, $r^*(y|x) = R(y), \forall x \in \Xepsalpc$. Here, $R(y)$ denotes any distribution over $\Xepsalpc$, i.e., $\sum_{y \in \Xepsalpc}R(y) = 1$, that is independent of $X$.
We call $r^*(y|x)$ the \emph{$X$-invariant randomization} \cite{Sadeghi2020ITW}, which incurs an $\alpha$-lift $\ell_{\alpha}(y) = \bar{\ell}_{\alpha}, \forall y \in \Xepsalpc$ where
\begin{equation}\label{eq:AlphaLiftYMerge}
    \bar{\ell}_{\alpha} \triangleq \Big( \sum_{s} p(s) \left( \frac{p(\Xepsalpc|s)}{p(\Xepsalpc)} \right)^{\alpha} \Big)^{1/\alpha}.
\end{equation}
Here, $p(\Xepsalpc|s) = \sum_{x \in \Xepsalpc} p(x|s)$ and $p(\Xepsalpc) = \sum_{x \in \Xepsalpc} p(x)$.
The following theorem shows that $\bar{\ell}_{\alpha}$ is a fundamental bound on $\ell_{\alpha}(y)$ resulted from the watchdog randomization $r(y|x)$.

\begin{theorem} \label{theo:main}
For all $\alpha \in (1,\infty)$ and $\eps>0$,
\begin{subequations} \label{eq:OptMechanism}
\begin{align}
    & \min_{r(y|x)} \max_{y \in \Xepsc} \ell_{\alpha} (y) = \bar{\ell}_{\alpha}, \label{eq:OptMechanismMax}\\
    & \min_{r(y|x)} \sum_{y \in \Xepsc} p(y) \ell_{\alpha}(y) = \sum_{y \in \Xepsc} p(y) \bar{\ell}_{\alpha} = p(\Xepsalpc) \bar{\ell}_{\alpha}. \label{eq:OptMechanismExp}
\end{align}
\end{subequations}
%\begin{enumerate}[(a)]
%  \item $\min_{p(y|x)} \max_{y \in \Xepsc} \ell_{\alpha} (y) = \bar{\ell}_{\alpha}$
%  \item $\min_{p(y|x)} \sum_{y \in \Xepsc} p(y) \ell_{\alpha} (y) = \bar{\ell}_{\alpha}$
%\end{enumerate}
\end{theorem}
\begin{IEEEproof}
Rewrite the $\alpha$-lift for all $y \in \Xepsalpc$ as
\[ \ell_{\alpha}(y) = \frac{\Big( \sum_{s} p(s) \Big( \sum_{x \in \Xepsc} r(y|x) p(x|s) \Big)^{\alpha} \Big)^{1/\alpha}}{\sum_{x \in \Xepsc}r(y|x) p(x)}  \]
and $\bar{\ell}_{\alpha} = \frac{\big( \sum_{s} p(s) p(\Xepsc|s)^{\alpha} \big)^{1/\alpha}}{p(\Xepsc)} $.
We prove \eqref{eq:OptMechanismMax} first.
Pick any $y \in \Xepsc$ and assume that for all $y' \in \Xepsc$ such that $y' \neq y$, $\ell_{\alpha}(y') < \bar{\ell}_{\alpha}$, i.e.,
\begin{equation*}
    p(\Xepsc) \Big( \sum_{s} p(s) p(y'|s)^{\alpha}\Big)^{1/\alpha} <  p(y') \Big( \sum_{s} p(s) p(\Xepsc|s)^{\alpha} \Big)^{1/\alpha}.
\end{equation*}
This necessarily makes $\ell_{\alpha}(y) > \bar{\ell}_{\alpha}$, i.e.,
\begin{equation} \label{eq:theo:mainIneq}
p(\Xepsc) \Big( \sum_{s} p(s) p(y|s)^{\alpha} \Big)^{1/\alpha} > p(y) \Big( \sum_{s} p(s) p(\Xepsc|s)^{\alpha} \Big)^{1/\alpha}.
\end{equation}
See Appendix~\ref{app:eq:theo:mainIneq} for the proof of \eqref{eq:theo:mainIneq}.
Therefore, we conclude that except the $X$-invariant randomization, there is no other randomization $r(y|x)$ that can incur an $\alpha$-lift $\ell_{\alpha}(y)$ such that $\ell_{\alpha}(y) < \bar{\ell}_{\alpha}$ for all $y \in \Xepsalpc$. This proves \eqref{eq:OptMechanismMax}.

Consider probabilities $p(y)$ and $p(y|s)$ for $y \in \Xepsalpc$. Recalling that $\sum_{y\in \Xepsalpc} r(y|x) = 1$, we have
\begin{align*}
     & \sum_{y \in \Xepsalpc} p(y) = \sum_{x, y \in \Xepsalpc}  r(y|x) p(x) % = \sum_{x\in \Xepsalpc}  (\sum_{y\in \Xepsalpc} r(y|x)) p(x)
     = p(\Xepsalpc), \\
     & \sum_{y \in \Xepsalpc} p(y|s) = \sum_{x, y \in \Xepsalpc} r(y|x) p(x|s) = p(\Xepsalpc|s).
\end{align*}
Consider $\sum_{y \in \Xepsalpc} p(y) \ell_{\alpha} (y) = \sum_{y \in \Xepsalpc} \left(  \sum_{s} p(s) p(y|s)^{\alpha} \right)^{1/\alpha}$. By Minkowski inequality,
\begin{multline*}
    \sum_{y \in \Xepsalpc} \Big(  \sum_{s} p(s) p (y|s)^{\alpha} \Big)^{1/\alpha} \geq  \Big(  \sum_{s} p(s) \Big( \sum_{y \in \Xepsalpc}p(y|s) \Big)^{\alpha} \Big)^{1/\alpha} \\
    = \Big(  \sum_{s} p(s) p(\Xepsalpc|s)^{\alpha} \Big)^{1/\alpha} = p(\Xepsalpc) \bar{\ell}_{\alpha} = \sum_{y \in \Xepsc} p(y) \bar{\ell}_{\alpha}
\end{multline*}
This proves \eqref{eq:OptMechanismExp}.
\end{IEEEproof}

Theorem~\ref{theo:main} clearly states that the $X$-invariant randomization minimizes the $\alpha$-lift.
It should be noted that the randomization $r(y|x)$, even if it is $X$-invariant, does not necessarily result in an $\alpha$-lift $\ell_{\alpha}(y)$ such that $ \log \ell_{\alpha}(y) \leq \eps$ for all $y \in \Xepsalpc$.
For a given value of $\eps' > 0$, we say that $(\eps',\alpha)$-log-lift is attainable in $\Xepsalpc$ if there exists a randomization $r(y|x)$ such that $\ell_{\alpha}(y) \leq \eps', \forall y \in \Xepsalpc$. The value $\bar{\ell}_{\alpha}$ can be used as the necessary and sufficient condition for the $(\eps',\alpha)$-log-lift attainability, which we state as the following corollary.

\begin{corollary} \label{coro:iffCond}
For all $\alpha \in (1,\infty)$, $(\eps',\alpha)$-log-lift is attainable in $\Xepsalpc$ iff $\eps' \geq \log \bar{\ell}_{\alpha}$.  \hfill\IEEEQED
\end{corollary}
By Theorem~\ref{theo:main}, the corollary below is also immediate.
\begin{corollary} \label{coro:Sibson}
For all $\alpha \in (1,\infty)$,
\begin{align*}
    \min_{r(y|x)} \bar{I}_{\alpha}^{S}(S;Y) =& \frac{\alpha}{\alpha-1} \log \max \Big\{ \max_{x \in \Xepsalp} \ell_{\alpha}(x), \bar{\ell}_{\alpha} \Big\}, \\
    \min_{r(y|x)} I_{\alpha}^{S}(S;Y) =& \frac{\alpha}{\alpha-1} \log \Big[ \sum_{x \in \Xepsalpc} p(x) \ell_{\alpha}(x) + p(\Xepsalpc) \bar{\ell}_{\alpha} \Big]. \quad\ \IEEEQED % \IEEEQEDhereeqn
\end{align*}
\end{corollary}
We omit the proofs of Corollaries~\ref{coro:iffCond} and \ref{coro:Sibson}, because the former is a direct result of Theorem~\ref{theo:main} and the latter is obtained simply by substituting \eqref{eq:OptMechanismMax} and \eqref{eq:OptMechanismExp} in $\bar{I}_{\alpha}^{S}(S;Y)$ and $I_{\alpha}^{S}(S;Y)$, respectively.

\section{Experiments} \label{sec:PUT}

 \begin{figure}[t]
   \centering
    \scalebox{0.7}{% This file was created by matlab2tikz v0.4.3.
% Copyright (c) 2008--2013, Nico Schlömer <nico.schloemer@gmail.com>
% All rights reserved.
%
% The latest updates can be retrieved from
%   http://www.mathworks.com/matlabcentral/fileexchange/22022-matlab2tikz
% where you can also make suggestions and rate matlab2tikz.
%
\begin{tikzpicture}

\begin{axis}[%
width=4.8in,
height=3.2in,
xmin=0,
xmax=1,
xlabel={\Large NMIL},
ymin=0,
ymax=1.25,
ylabel={\Large privacy leakage},
]
\addplot [
color=blue,
dashed,
mark=asterisk,
line width = 1,
mark options={solid},
]
table[row sep=crcr]{
1.31153990413789e-16 0.449075648646785\\
0.00857508902924729 0.311004954725512\\
0.0208029920717376 0.309700628191654\\
0.0382382679515912 0.296080503247326\\
0.0599501359229249 0.28844379262252\\
0.0821937621542223 0.280876619505707\\
0.11046702956072 0.267978094045835\\
0.141083244321407 0.263047048031002\\
0.17027407881809 0.26248802062488\\
0.203273979482015 0.23851579080286\\
0.232443179864163 0.228987490806336\\
0.266138017471941 0.224263482652533\\
0.298617884849402 0.220801602166193\\
0.331045576330531 0.214450897329921\\
0.368472286111642 0.208117566642452\\
0.402561412991841 0.205372809644225\\
0.442122403811151 0.205074084668732\\
0.473568792090571 0.205029862829245\\
0.507254529760097 0.196496516549499\\
0.546808859725192 0.187066872263017\\
0.587278771240759 0.176000662409608\\
0.630726170567803 0.160830940137197\\
0.670037811306403 0.147135147688247\\
0.717684601700249 0.132724361112919\\
0.761916574729866 0.12938760650232\\
0.804567681682232 0.127768200971838\\
0.851840043068621 0.113582141405168\\
0.898687773854185 0.110026293796324\\
0.948147527020141 0.0927645605190504\\
1 0\\
};
\addlegendentry{$\min_{r(y|x)} \bar{I}_{\alpha}^{S}(S;Y)$ for $\alpha = 1.5$};

\addplot [
color=red,
dashed,
mark=o,
mark options={solid},
line width = 1,
]
table[row sep=crcr]{
1.31153990413789e-16 0.21142315186318\\
0.00857508902924729 0.202446319978714\\
0.0208029920717376 0.194886426104485\\
0.0382382679515912 0.18774433615809\\
0.0599501359229249 0.175102370268171\\
0.0821937621542223 0.163633706416414\\
0.11046702956072 0.152963663738728\\
0.141083244321407 0.145768262389764\\
0.17027407881809 0.136706508418108\\
0.203273979482015 0.127830910947945\\
0.232443179864163 0.121689790428074\\
0.266138017471941 0.110649832555533\\
0.298617884849402 0.103726856093214\\
0.331045576330531 0.09770702445786\\
0.368472286111642 0.0893538928783431\\
0.402561412991841 0.0822749418003442\\
0.442122403811151 0.0739969842444649\\
0.473568792090571 0.0674112360534125\\
0.507254529760097 0.0619013173131888\\
0.546808859725192 0.0552075862454793\\
0.587278771240759 0.047996594006281\\
0.630726170567803 0.0404056668022445\\
0.670037811306403 0.0347536507040327\\
0.717684601700249 0.0284528434819469\\
0.761916574729866 0.0232177440957821\\
0.804567681682232 0.0186558161057516\\
0.851840043068621 0.0134125320861705\\
0.898687773854185 0.00887005672246008\\
0.948147527020141 0.00415745508751766\\
1 0\\
};
\addlegendentry{$\min_{r(y|x)} I_{\alpha}^{S}(S;Y)$ for $\alpha = 1.5$};

\addplot [
color=green,
line width = 1,
dashed,
mark=square,
mark options={solid},
]
table[row sep=crcr]{
1.31153990413789e-16 0.728029918401099\\
0.00800870882677662 0.678123186515593\\
0.0208029920717376 0.670097558334657\\
0.0379935463675719 0.573249442159845\\
0.0596414483845599 0.545092091902154\\
0.0822314159217331 0.540259071724347\\
0.10793333712201 0.527890539293478\\
0.133277853731134 0.503390219866851\\
0.159397663649672 0.481846229966379\\
0.187764521866605 0.480146694343941\\
0.217381552514184 0.47860457012167\\
0.24835547691741 0.450578652283619\\
0.282531953213427 0.439796936950922\\
0.318616439319365 0.42790472185078\\
0.356105993788778 0.427075235138893\\
0.392730029673243 0.426905853366169\\
0.43202283604157 0.425669069434292\\
0.473012044906731 0.403481320827184\\
0.512383239571932 0.401802917528621\\
0.54465518010651 0.388649812764937\\
0.582882644542974 0.378912210161579\\
0.625358899612756 0.378765930936161\\
0.669321740294577 0.37188576248828\\
0.713680020575581 0.34241438597885\\
0.761916574729866 0.327105561228145\\
0.807798164941298 0.308826861823462\\
0.855117283758707 0.279450155514727\\
0.898687773854185 0.225692729969071\\
0.948147527020141 0.217934230635404\\
1 0\\
};
\addlegendentry{$\min_{r(y|x)} \bar{I}_{\alpha}^{S}(S;Y)$ for $\alpha = 10$};

\addplot [
color=orange,
dashed,
mark=triangle,
mark options={solid},
line width = 1,
]
table[row sep=crcr]{
1.31153990413789e-16 0.439662439854648\\
0.00800870882677662 0.435208864559344\\
0.0208029920717376 0.40905003719679\\
0.0379935463675719 0.394699961684927\\
0.0596414483845599 0.381920394145824\\
0.0822314159217331 0.370133895564615\\
0.10793333712201 0.349162529485941\\
0.133277853731134 0.336836703216963\\
0.159397663649672 0.327104774790879\\
0.187764521866605 0.315978128335427\\
0.217381552514184 0.304359710091571\\
0.24835547691741 0.278588076023482\\
0.282531953213427 0.259295992119648\\
0.318616439319365 0.247877980522948\\
0.356105993788778 0.229248680357053\\
0.392730029673243 0.207592490493911\\
0.43202283604157 0.192926293134446\\
0.473012044906731 0.17994748654325\\
0.512383239571932 0.161216550293111\\
0.54465518010651 0.145879335971956\\
0.582882644542974 0.130598783125801\\
0.625358899612756 0.112493997636113\\
0.669321740294577 0.0964123069600072\\
0.713680020575581 0.0779328461959509\\
0.761916574729866 0.0624456046719378\\
0.807798164941298 0.0501454127360214\\
0.855117283758707 0.0351487795269769\\
0.898687773854185 0.0226092683694998\\
0.948147527020141 0.0112901480936431\\
1 0\\
};
\addlegendentry{$\min_{r(y|x)} I_{\alpha}^{S}(S;Y)$ for $\alpha = 10$};

\addplot [
color=purple,
dashed,
mark=diamond,
mark options={solid},
line width = 1,
]
table[row sep=crcr]{
0 0.865152305702695\\
0.00800870882677662 0.848420253986159\\
0.0218321686700191 0.773043462487336\\
0.0379935463675719 0.739525143658701\\
0.0596414483845599 0.672404924383298\\
0.0826874455902929 0.668818729968512\\
0.106260534144967 0.66709757050469\\
0.133474994996364 0.631596345115327\\
0.160476106364451 0.621669785511002\\
0.188901954133276 0.615079603438547\\
0.217381552514184 0.598147394485182\\
0.252800155454614 0.595162430388755\\
0.288657642743059 0.587420073665414\\
0.324011231935581 0.577859747123304\\
0.357692168219186 0.567971095137431\\
0.394920590679345 0.56223483077424\\
0.433635686756293 0.55850151663782\\
0.471749326078954 0.550802009684818\\
0.503532639889588 0.523043631294666\\
0.544818279468513 0.516223743285001\\
0.586587331831093 0.500483945999996\\
0.630021177327684 0.48886578748091\\
0.669321740294577 0.470349187320914\\
0.714065849067805 0.459528845732114\\
0.758980322054228 0.446008489263315\\
0.807798164941298 0.413379747945533\\
0.85091331130071 0.404883109805854\\
0.898687773854185 0.368571018012919\\
0.95001183415561 0.335323565608092\\
1 2.22044604925031e-16\\
};
\addlegendentry{$\min_{r(y|x)} \bar{I}_{\infty}^{S}(S;Y)$};

\end{axis}
\end{tikzpicture}% 
}
	\caption{Privacy-utility tradeoff: the reduction of $I_{\alpha}^{S}(S;Y)$ and $\bar{I}_{\alpha}^{S}(S;Y)$ as the data utility, measured by the normalized mutual information loss (NMIL) decreases, by the randomization scheme in~\eqref{eq:Mechanism} that is $X$-invariant. }\bigskip
	\label{fig:PUT}
 \end{figure}
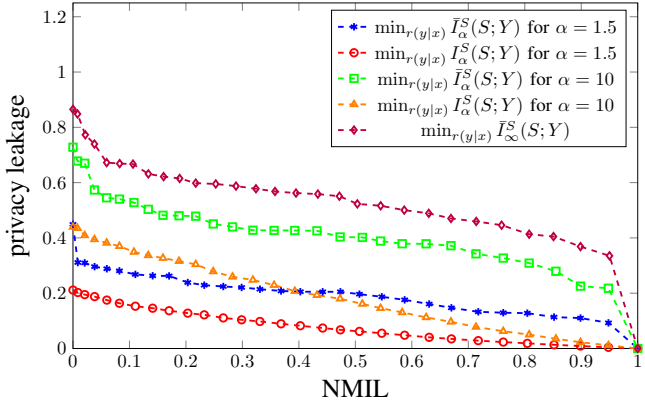

For the bi-partition $\Set{\Xeps, \Xepsc}$, the data utility under the $X$-invariant randomization $r^*(y|x)$ can be measured by the mutual information
$I(X;Y) = - \sum_{x \in \Xepsalp} p(x) \log p(x) - p(\Xepsalpc) \log p(\Xepsalpc)$.
Then, $H(X) - I(X;Y)$ denotes the loss in data utility. In this section, we use the normalized mutual information loss (NMIL)
\begin{equation}
    \text{NMIL}\triangleq \frac{H(X) - I(X;Y)}{H(X)}.
\end{equation}
%
%\PS{For a given joint distribution $p_{S,X}$ and $\alpha$, to obtain the} bi-partition $\Set{\Xeps, \Xepsc}$, a smaller value of $\eps$ indicates that a more strict data privacy constraint is applied. In this case, \PS{the `high risk'} elements in $\Xepsc$ not satisfying $(\eps,\alpha)$-log-lift \PS{need} to be randomized.
%%
%Under the $X$-invariant randomization, the data utility can be measured by the mutual information
%\begin{equation}
%    I(X;Y) = - \sum_{x \in \Xepsalp} p(x) \log p(x) - p(\Xepsalpc) \log p(\Xepsalpc).
%\end{equation}
%\PS{Therefore, starting from $\eps = \infty$, $\Xepsalp = \X$ and $\Xepsalpc = \emptyset$, where the utility is maximum and equal to $I(X;Y) = H(X)$, as $\eps$ decreases to enforce more data privacy, more elements will be added to $\Xepsalpc$, losing the fidelity of the released data $Y$. In this section, we measure the loss in utility by normalized mutual information loss (NMIL), defined as
%\begin{align}
%    \text{NMIL}&\triangleq \frac{H(X) - I(X;Y)}{H(X)} \\&= \frac{p(\Xepsalpc) \log p(\Xepsalpc)-\sum_{x \in \Xepsalpc} p(x) \log p(x)}{H(X)}
%\end{align}
%}
%%It is clear that $I(X;Y)$ is decreasing in $|\Xepsc|$.
%%
%%Therefore, as $\eps$ decreases, we are enforcing data privacy, but losing the fidelity of the released data $Y$.
It is clear that a smaller value of $\eps$ indicates a more strict data privacy constraint, and therefore, more `high risk' elements in $\Xepsc$ are not satisfying $(\eps,\alpha)$-log-lift. This necessarily increases the $\text{NMIL}$. That is, as $\eps$ decreases, we are enforcing data privacy, but losing the fidelity of the released data $Y$.
For a countable alphabet $\X$, we only need to focus on a finite set of values for $\eps$.
Given any $\alpha \in (1,\infty)$, let $\Phi = (x_1,\dotsc,x_{|\X|})$ be the ordering of elements in $\X$ such that $x_i \neq x_{i'},\forall i \neq i'$ and $\ell_{\alpha}(x_i) > \ell_{\alpha}(x_{i+1}), \forall i \in \{1,\dotsc,|\X|-1\}$.\footnote{We assume the values of $\ell_{\alpha}(x)$ for all $x \in \X$ are different. If there are identical $\alpha$-lifts, one can obtain the set $\tilde{\X} \subsetneq \X$ containing the unique values of $\alpha$-lift and order $\ell_{\alpha}(x)$ on $\tilde{\X}$. }
Then, setting $\Xepsc = (x_1, \dotsc, x_i)$ is equivalent to applying $\eps = \ell_{\alpha}(x_{i+1})$  to obtain the cut $\Set{\Xeps, \Xepsc}$. The privacy-utility tradeoff (PUT) can be observed by increasing $i$ from $1$ to $|\X|$.

We randomly generated a joint probability distribution $p(s,x)$ with $|\S| = 20$ and $|\X| = 30$. For all $i \in \Set{1,\dotsc,|\X|}$, we obtained $\Xepsc = (x_1, \dotsc, x_i)$ and the resulting $\min_{r(y|x)} {I}_{\alpha}^{S}(S;Y)$ and $\min_{r(y|x)} \bar{I}_{\alpha}^{S}(S;Y)$ by using an optimal $X$-invariant randomization. Fig.~\ref{fig:PUT} shows these two privacy measures vs. the NMIL.
For $\min_{r(y|x)} \bar{I}_{\alpha}^{S}(S;Y)$, it can be seen that as $\alpha$ increases, the PUT converges to $\min_{r(y|x)} \bar{I}_{\infty}^{S}(S;Y)$.
Comparing the PUT for $\min_{r(y|x)} \bar{I}_{1.5}^{S}(S;Y)$ to $\min_{r(y|x)} I_{10}^{S}(S;Y)$, an interesting finding is that increasing $\alpha$ but relaxing $\max_{y \in \Y} \ell_\alpha(y)$ to $\E[\ell_\alpha(Y)]$ can improve the data utility. It will be worth studying how the effect of the maximum lift over $\X$ differs from that over $\S$.\footnote{Recall that as $\alpha$ increase, $\ell_{\alpha}(x) \rightarrow \max_{s \in \S} l(s,x)$, i.e., we become more conservative about the `high risk' events along the $S$ dimension. But, the conservativeness over the $X$ dimension is tuned by $I_{\alpha}^{S}(S;Y)$ and $\bar{I}_{\alpha}^{S}(S;Y)$: $I_{\alpha}^{S}(S;Y) = \frac{\alpha}{\alpha-1} \log \E[\ell_{\alpha}(Y)]$ measures the expected value of $\ell_{\alpha}(y)$, while $\bar{I}_{\alpha}^{S}(S;Y) = \frac{\alpha}{\alpha-1} \log \max_{y \in \Y}\ell_{\alpha}(Y)$ focuses on the maximum of $\ell_{\alpha}(y)$.}

\subsection{Comparison with Existing Privacy Watchdog Methods}

For the watchdog method in \cite{Watchdog2019,Sadeghi2020ITW}, the privatized randomization $r(y|x)$ is applied to the set
\begin{equation} \label{eq:Xepscbar}
\Xepscbar = \Set{ x \in \X \colon \max_{s \in \S} |i(s,x)| > \bar\eps}.
\end{equation}
Here, $\bar{\eps} > 0$ denotes the threshold that discriminates `high risk' elements in $\X$ based on the maximum absolute value of the log-lift $\max_{s \in \S} |i(s,x)| = \max_{s \in \S} |\log l(s,x)|$.
We use the $\alpha$-lift to relax this watchdog method as follows.

Choosing $\alpha \in (1,\infty)$ and $\eps < \bar{\eps}$, we impose another constraint on the $\alpha$-lift and apply randomization on the set
\begin{equation} \label{eq:Xdepscbar}
\begin{aligned}
\Xdepscbar &= \Set{ x \in \Xepscbar \colon \log \ell_\alpha(x) > \eps} \\
&= \Set{ x \in \X \colon \log \ell_\alpha(x) > \eps, \max_{s \in \S} |i(s,x)| > \bar\eps}.
\end{aligned}
\end{equation}
In other words, we expand the privacy criterion: an element $x \in \X$ must have both $\ell_\alpha(x)$ and $\max_{s \in \S} |i(s,x)|$  exceed respective thresholds $\eps$ and $\bar{\eps}$  to be considered `high risk'.
We necessarily have $\Xdepscbar \subseteq \Xepscbar$, i.e., fewer elements in $\X$ are perturbed, and therefore the data utility is improved.
This method sacrifices data privacy to enhance the utility, where the tradeoff is controlled by the constraint $\log \ell_\alpha(x) > \eps$.

A similar, but somewhat composite approach is adopted in \cite[Section~V-A]{Sadeghi2020ITW}, where the set $\Xepscbar$ is refined by $\delta > 0$, a threshold on the probability for the resulting lift value to be $|i(s,y)| > \bar\eps$.
The elements $x \in \Xepscbar$ are iteratively removed from $\Xepscbar$ if
\vspace{-10pt}
\begin{multline} \label{eq:delta}
    \sum_{x' \in \Xepsbar \cup \Set{x}, s \in \S \colon |i(s,x')| > \bar{\eps}} p(s,x') \\
    + \sum_{s \in \S \colon |i(s,\Xepscbar \setminus \Set{x})|>\bar{\eps}} p(s,\Xepscbar \setminus \Set{x}) \leq \delta,
\end{multline}
where $i(s,\Xepscbar \setminus \Set{x}) = \log \frac{p(\Xepscbar \setminus \Set{x} | s)}{p(\Xepscbar \setminus \Set{x})}$ and $p(s,\Xepscbar\setminus \Set{x}) = \sum_{x' \in \Xepscbar \setminus \Set{x}} p(s,x')$.
Here, the cut $\Set{\Xepsbar \cup \Set{x}, \Xepsbar \setminus \Set{x}}$ denotes that element $x$ is removed from the `high risk' set $\Xepsbar$ and the LHS of \eqref{eq:delta} measures the probability that the resulting maximum absolute value of the log-lift exceeds $\bar\eps$ if the $X$-invariant randomization is applied to $\Xepsbar \setminus \Set{x}$.

 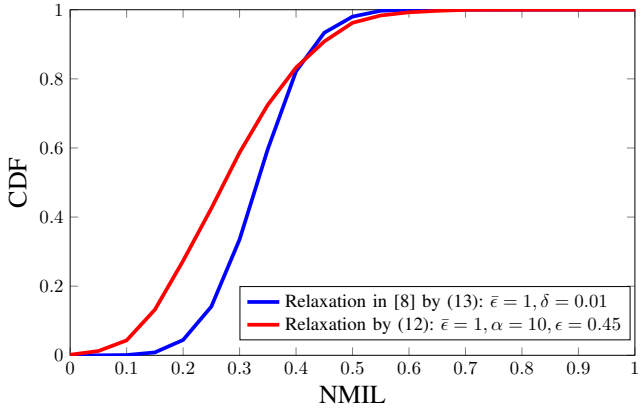
\begin{figure}[t]
	\centering
    \scalebox{0.7}{% This file was created by matlab2tikz v0.4.3.
% Copyright (c) 2008--2013, Nico Schlömer <nico.schloemer@gmail.com>
% All rights reserved.
%
% The latest updates can be retrieved from
%   http://www.mathworks.com/matlabcentral/fileexchange/22022-matlab2tikz
% where you can also make suggestions and rate matlab2tikz.
%

%
\begin{tikzpicture}

\begin{axis}[%
width=4.8in,
height=3.2in,
xmin=0,
xmax=1,
xlabel={\Large NMIL},
ymin=0,
ymax=1,
ylabel={\Large  CDF},
legend style={at={(0.3,0.20)},anchor=north west,draw=black,fill=white,legend cell align=left}
]

\addplot [
color=blue,
solid,
line width=2.0pt,
]
table[row sep=crcr]{
0 0\\
0.05 0\\
0.1 0.0008\\
0.15 0.0084\\
0.2 0.0444\\
0.25 0.141\\
0.3 0.3352\\
0.35 0.5968\\
0.4 0.8218\\
0.45 0.9334\\
0.5 0.98\\
0.55 0.997\\
0.6 0.9996\\
0.65 1\\
0.7 1\\
0.75 1\\
0.8 1\\
0.85 1\\
0.9 1\\
0.95 1\\
1 1\\
};
\addlegendentry{Relaxation in \cite{Sadeghi2020ITW} by \eqref{eq:delta}: $\bar{\eps}=1, \delta = 0.01$ };

\addplot [
color=red,
solid,
line width=2.0pt,
]
table[row sep=crcr]{
0 0.0016\\
0.05 0.0124\\
0.1 0.0432\\
0.15 0.132\\
0.2 0.2734\\
0.25 0.4252\\
0.3 0.5866\\
0.35 0.725\\
0.4 0.8326\\
0.45 0.9082\\
0.5 0.9622\\
0.55 0.9834\\
0.6 0.9926\\
0.65 0.9968\\
0.7 0.9996\\
0.75 0.9998\\
0.8 1\\
0.85 1\\
0.9 1\\
0.95 1\\
1 1\\
};
\addlegendentry{Relaxation by \eqref{eq:Xdepscbar}: $\bar{\eps} = 1, \alpha = 10, \eps = 0.45$ };

\end{axis}
\end{tikzpicture}% 
}
%	\caption{The CDF of NMIL comparing between two relaxation methods for the original watchdog method in \cite{Watchdog2019,Sadeghi2020ITW}: the set $\Xdepscbar$ in \eqref{eq:Xdepscbar} using the $\alpha$-lift; the refinement of $\Xepscbar$ using the condition \eqref{eq:delta} proposed in \cite[Section~V-A]{Sadeghi2020ITW}. }
\caption{The CDF of NMIL comparing the relaxation method in this paper (using  $\alpha$-lift to obtain $\Xdepscbar$ from \eqref{eq:Xdepscbar}) versus the refinement of $\Xepscbar$ using the condition \eqref{eq:delta} proposed in \cite[Section~V-A]{Sadeghi2020ITW}.}
	\label{fig:ITW_ISIT} \medskip
\end{figure}

We do the following experiment to compare the two relaxation methods.
Set $|\S| = 15$ and $|\X| = 20$ and $\bar{\eps} = 1$. Uniformly randomly generate the joint probability distribution $p(s,x)$. To obtain the relaxation set $\Xdepscbar$ in \eqref{eq:Xdepscbar} based on the $\alpha$-lift, we set $\alpha = 10$, $\eps = 0.45$ and apply the $X$-invariant randomization $r^*(y|x)$ to $\Xdepscbar$; For the relaxation method proposed in \cite[Section~V-A]{Sadeghi2020ITW}, we first obtain $\Xepscbar$ and then set $\delta = 0.01$, refine the set $\Xepscbar$ by \eqref{eq:delta} and apply $r^*(y|x)$.\footnote{As in \cite[Algorithm~1]{Sadeghi2020ITW}, an additional parameter $\eps_{\text{max}} = 4$ is applied to control the resulting maximum absolute value of the log-lift. That is, an $x \in \Xepscbar$ is removed from $\Xepscbar$ if \eqref{eq:delta} holds and $ \max\Set{\max_{x' \in \Xepsbar \cup \Set{x} |i(s,x')|, i(s,\Xepscbar \setminus \Set{x}) }} \leq \eps_{\text{max}}$. The blue plot is the same as \cite[Fig.~2]{Sadeghi2020ITW}. }
Repeating this procedure for $5000$ times, we plot the cumulative distribution function (CDF) of the resulting NMIL in Fig.~\ref{fig:ITW_ISIT}.
By and large, applying $X$-invariant randomization $r^*(y|x)$ to the refined subset $\Xdepscbar$ produces lower NMIL more frequently, but incurs a slightly larger tail for high NMIL. It can produce better utility (smaller NMIL) compared to the relaxation method proposed in~\cite[Section~V-A]{Sadeghi2020ITW}.
We also measure the resulting $\delta$, the probability that the maximum absolute value of the log-lift goes above $\bar{\eps} = 1$, for the relaxation method based on $\Xdepscbar$. We find that for 99\% of the time, $\delta < 0.024$.
Finally, the relaxation method in~\cite[Section~V-A]{Sadeghi2020ITW} is a two-step approach: $\Xepscbar$ is obtained first and then recursively refined by \eqref{eq:delta}, while the set $\Xdepscbar$ can be obtained in one go (once $\alpha$ and $\eps$ are set). That is, the relaxation by the $\alpha$-lift is less complex than \cite[Section~V-A]{Sadeghi2020ITW}.

\section{Conclusion}

We proposed an $\alpha$-lift $\ell_{\alpha}(x)$ for each instance $x \in \X$, which is tunable between the existing maximum lift (over $\S$) and the weighted lift w.r.t. the marginal probability $p(s)$ of the sensitive data $S$. We adopted the $\alpha$ privacy watchdog privatization scheme, which is the existing watchdog method based on $\alpha$-lift, and studied its optimality. A threshold $\eps$ is applied to filter out $\Xepsalpc$ containing all $x$'s such that the logarithm of $\alpha$-lift exceeds $\eps$. While the set $\X \setminus \Xepsalpc$ is released unperturbed, we proved that the optimal randomization scheme $r^*(y|x), \forall x,y \in \Xepsalpc$ is $X$-invariant.
We revealed the interpretation of the Sibson mutual information $I_{\alpha}^{S}(S ; Y)$ in data privacy in terms of $\E[\ell_{\alpha}(y)]$ and defined $\bar{I}_{\alpha}^{S}(S ; Y)$ using $\max_{y \in \Y} \ell_{\alpha}(y)$. It was shown that an $X$-invariant $r^*(y|x)$ also minimizes $I_{\alpha}^{S}(S ; Y)$ and $\bar{I}_{\alpha}^{S}(S ; Y)$.

There are some aspects that can be further studied following this paper.
First, the interpretation of the Sibson mutual information in data privacy should be further explored by observing the role of $\alpha$ when it changes from $1$ to $\infty$. The study should also involve a comparison to the Arimoto mutual information that is used in the $\alpha$-leakage in \cite{Liao2019Alpha}.
Second, for $l(s,y) = \frac{p(s|y)}{p(s)}$ denoting how different $p(s|y)$ is from $p(s)$, a smaller value of $\min_{s \in \S} l(s,y)$ also denotes a high privacy leakage. Therefore, it is worth exploring how one can propose an $\alpha$-lift that is tunable between not only $\max_{s \in \S} l(s,y)$, but also $\min_{s \in \S} l(s,y)$.
Third, as pointed out in Section~\ref{sec:PUT}, it worth understanding the tradeoff between taking maximum $\alpha$-lift over $\Y$ versus increasing $\alpha$.

%%%%%%
%% Appendix:
%% If needed a single appendix is created by
%%
%\appendix
%%
%% If several appendices are needed, then the command
%%
\appendices
%%
%% in combination with further \section-commands can be used.
%%%%%%

\section{Example for $\alpha$-lift} \label{app:example}
\begin{figure}[t]
   \centering
    \scalebox{0.6}{% This file was created by matlab2tikz v0.4.3.
% Copyright (c) 2008--2013, Nico Schlömer <nico.schloemer@gmail.com>
% All rights reserved.
%
% The latest updates can be retrieved from
%   http://www.mathworks.com/matlabcentral/fileexchange/22022-matlab2tikz
% where you can also make suggestions and rate matlab2tikz.
%
\begin{tikzpicture}

\begin{axis}[%
width=4.5in,
height=3.2in,
view={-37.5}{30},
scale only axis,
xmin=1,
xmax=4,
xtick={1,2,3,4},
xticklabels={$a$,$b$,$c$,$d$},
xlabel={\Large $X$},
xmajorgrids,
ymin=0,
ymax=35,
ytick={1,5,9,13,17,21,25,29,33},
yticklabels={0.05,0.15,0.25,0.35,0.45,0.55,0.65,0.75,0.85},
ylabel={\Large $\rho$},
ymajorgrids,
zmin=1,
zmax=1.4,
zlabel={\Large $\ell_\alpha(x)$},
zmajorgrids,
]

\addplot3[%
surf,
shader=faceted,
draw=black,
mesh/rows=4
]
table[row sep=crcr,header=false] {
1 1 1.01123298611139\\
1 2 1.01693871903084\\
1 3 1.02270150451974\\
1 4 1.02851895445316\\
1 5 1.03438815139029\\
1 6 1.04030556619549\\
1 7 1.04626696221042\\
1 8 1.05226728352931\\
1 9 1.05830052442584\\
1 10 1.06435957635962\\
1 11 1.07043604822209\\
1 12 1.07652005452541\\
1 13 1.0825999650425\\
1 14 1.08866210790363\\
1 15 1.09469041625384\\
1 16 1.10066600615807\\
1 17 1.10656667034498\\
1 18 1.1123662683894\\
1 19 1.11803398874989\\
1 20 1.12353345129265\\
1 21 1.12882160997512\\
1 22 1.13384740342719\\
1 23 1.13855008510662\\
1 24 1.14285714285714\\
1 25 1.14668168762458\\
1 26 1.14991914915214\\
1 27 1.15244305716161\\
1 28 1.15409960129021\\
1 29 1.15470053837925\\
1 30 1.15401382970217\\
1 31 1.15175110689979\\
1 32 1.14755062109849\\
1 33 1.14095361339933\\
1 34 1.13137084989848\\
1 35 1.11803398874989\\
2 1 1.00622650617172\\
2 2 1.00931736247924\\
2 3 1.01238863482481\\
2 4 1.01543641411519\\
2 5 1.01845641521172\\
2 6 1.02144393875656\\
2 7 1.0243938285881\\
2 8 1.02730042415834\\
2 9 1.03015750727543\\
2 10 1.03295824238872\\
2 11 1.03569510950935\\
2 12 1.03835982871201\\
2 13 1.04094327498895\\
2 14 1.04343538201927\\
2 15 1.04582503316759\\
2 16 1.0480999377282\\
2 17 1.05024649007095\\
2 18 1.05224960891024\\
2 19 1.05409255338946\\
2 20 1.05575671202694\\
2 21 1.05722135977627\\
2 22 1.05846337747357\\
2 23 1.05945692672795\\
2 24 1.06017307179005\\
2 25 1.06057933802047\\
2 26 1.06063919415511\\
2 27 1.06031144246848\\
2 28 1.05954949694951\\
2 29 1.05830052442584\\
2 30 1.05650441677917\\
2 31 1.05409255338946\\
2 32 1.05098630087051\\
2 33 1.04709518076455\\
2 34 1.0423146132941\\
2 35 1.03652311372649\\
3 1 1.41839145496814\\
3 2 1.47914555789843\\
3 3 1.50519932234904\\
3 4 1.51185789203691\\
3 5 1.50713905929225\\
3 6 1.49558143578106\\
3 7 1.47989278146361\\
3 8 1.46175013083555\\
3 9 1.4422205101856\\
3 10 1.42199574954089\\
3 11 1.40152977645347\\
3 12 1.38112195169243\\
3 13 1.36096923288574\\
3 14 1.34119967493559\\
3 15 1.32189441495554\\
3 16 1.30310236689807\\
3 17 1.2848501971897\\
3 18 1.2671491862494\\
3 19 1.25\\
3 20 1.23339603778985\\
3 21 1.21732579798162\\
3 22 1.20177455794346\\
3 23 1.1867255707788\\
3 24 1.17216091849053\\
3 25 1.15806211911111\\
3 26 1.14441055657821\\
3 27 1.13118778229637\\
3 28 1.11837572348797\\
3 29 1.10595682369059\\
3 30 1.09391413383097\\
3 31 1.08223136734312\\
3 32 1.07089292921506\\
3 33 1.05988392624555\\
3 34 1.04919016388832\\
3 35 1.03879813366217\\
4 1 1.01431813972814\\
4 2 1.0216720940756\\
4 3 1.02915852781945\\
4 4 1.03677906057744\\
4 5 1.04453499818249\\
4 6 1.05242725455596\\
4 7 1.06045625748941\\
4 8 1.06862183485353\\
4 9 1.07692307692308\\
4 10 1.08535816945592\\
4 11 1.09392419082666\\
4 12 1.10261686480383\\
4 13 1.11143025835627\\
4 14 1.12035641101958\\
4 15 1.12938487863156\\
4 16 1.13850216935696\\
4 17 1.14769104345249\\
4 18 1.15692963959349\\
4 19 1.16619037896906\\
4 20 1.17543858257643\\
4 21 1.18463071549575\\
4 22 1.19371214189013\\
4 23 1.20261423230209\\
4 24 1.211250604822\\
4 25 1.21951219512195\\
4 26 1.22726072339654\\
4 27 1.23431993679564\\
4 28 1.24046371755307\\
4 29 1.24539969815448\\
4 30 1.24874630825981\\
4 31 1.25\\
4 32 1.24848739440392\\
4 33 1.24329354326344\\
4 34 1.23315090602278\\
4 35 1.2162606385263\\
};
\end{axis}
\end{tikzpicture}% 
}
	\caption{The $\alpha$-lift $\ell_{\alpha}(x)$ for the conditional probability $p(x|s)$ in Table~\ref{tab:Example}. Here, $\rho = p(S = 1)$. }
	\label{fig:ellExample}
 \end{figure}
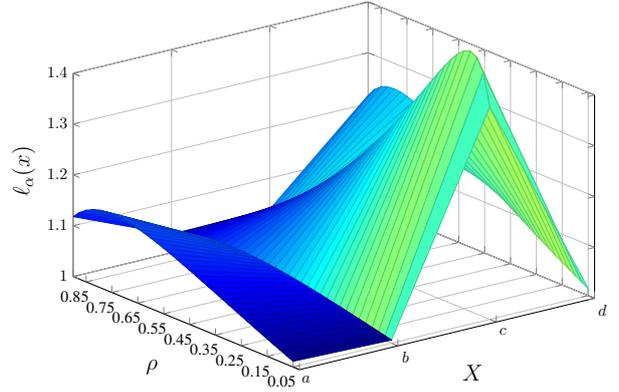

Rewrite the $\alpha$-lift \eqref{eq:AlphaLiftX} as $\ell_{\alpha}(x) =\frac{(\sum_{s} p(s) p(x|s)^\alpha)^{1/\alpha}}{\sum_{s} p(s)p(x|s)}$. This expression states that increasing the marginal probability $p(s)$ for $s$ having larger $p(x|s)$ does not necessarily increase the value of $\ell_{\alpha}(x)$.
For the conditional probability $p(x|s)$ in Table~\ref{tab:Example}, consider the marginal probability $p(S = 1) = \rho$ and $p(S = 2) = 1- \rho$ for $\rho \in [0,1]$. For $\alpha = 2$, consider the instance $X = c$, where $p(X = c |S = 1) > p(X = c |S = 2)$.
The $\alpha$-lift is $\ell_{2}(X = c) = \frac{\sqrt{0.01+0.48 \rho}}{0.1+0.6 \rho}$. Observe the change of $\ell_{2}(X = c)$ as $\rho$ increases. For $\rho \leq 0.125$, $\frac{d \ell_{2}(X = c)}{d \rho} \geq 0$ and so $\ell_{2}(X = c)$ is increasing. But, for $\rho > 0.125$, $\frac{d \ell_{2}(X = c)}{d \rho} < 0$, $\ell_{2}(X = c)$ is decreasing. See Fig.~\ref{fig:ellExample}.

% \vspace{10pt}

\section{Proof of \eqref{eq:theo:mainIneq}} \label{app:eq:theo:mainIneq}
\vspace{-10pt}
\begin{align}
    & p(\Xepsc) \Big( \sum_{s} p(s) p(y|s)^{\alpha} \Big)^{1/\alpha} - p(y) \Big( \sum_{s} p(s) p(\Xepsc|s)^{\alpha} \Big)^{1/\alpha} \nonumber \\
    &= p(\Xepsc) \Big( \sum_{s} p(s) \big( \sum_{x \in \Xepsc} r(y|x)p(x|s) \big)^{\alpha} \Big)^{1/\alpha} \nonumber \\
    & \quad        - \Big( \sum_{x \in \Xepsc} r(y|x) p(x) \Big)  \Big( \sum_{s} p(s) p(\Xepsc|s)^{\alpha} \Big)^{1/\alpha} \nonumber \\
%    &= p(\Xepsc) \Big( \sum_{s} p(s) \big( \sum_{x \in \Xepsc} \big( 1- \sum_{y' \in \Xepsc \colon y' \neq y}r(y'|x) \big) p(x|s) \big)^{\alpha} \Big)^{\frac{1}{\alpha}} \nonumber  \\
%    & - \Big( \sum_{x \in \Xepsc} \big( 1- \sum_{y' \in \Xepsc \colon y' \neq y}r(y'|x) \big) p(x) \Big) \Big( \sum_{s} p(s) p^{\alpha}(\Xepsc|s) \Big)^{\frac{1}{\alpha}} \nonumber \\
    &= p(\Xepsc) \Big( \sum_{s} p(s) \Big( p(\Xepsc|s) - \sum_{y' \in \Xepsc \colon y' \neq y} p(y'|s) \Big)^{\alpha} \Big)^{1/\alpha} - \nonumber  \\
    & \quad \Big( p(\Xepsc) - \sum_{y' \in \Xepsc \colon y' \neq y} p(y') \Big) \Big( \sum_{s} p(s) p(\Xepsc|s)^{\alpha} \Big)^{1/\alpha} \label{eq:MinkEq1} \\
    &\geq  p(\Xepsc) \Big[ \Big(\sum_{s} p(s) p(\Xepsc|s)^{\alpha} \Big)^{1/\alpha}  -  \nonumber \\
    & \quad \Big( \sum_{s} p(s)\Big( \sum_{y' \in \Xepsc \colon y' \neq y}  p(y'|s) \Big)^{\alpha} \Big)^{1/\alpha} \Big] - \nonumber  \\
    & \quad \Big( p(\Xepsc) - \sum_{y' \in \Xepsc \colon y' \neq y} p(y') \Big) \Big( \sum_{s} p(s) p(\Xepsc|s)^{\alpha} \Big)^{1/\alpha} \label{eq:MinkIneq1}\\
    &= \sum_{y' \in \Xepsc \colon y' \neq y} p(y') \Big( \sum_{s} p(s) p(\Xepsc|s)^{\alpha} \Big)^{\frac{1}{\alpha}} - \nonumber \\
    &\qquad p(\Xepsc) \Big( \sum_{s} p(s)\Big( \sum_{y' \in \Xepsc \colon y' \neq y} p(y'|s) \Big)^{\alpha} \Big)^{\frac{1}{\alpha}} \nonumber\\
    &\geq  \sum_{y' \in \Xepsc \colon y' \neq y} \Big[p(y')  \Big( \sum_{s} p(s) p(\Xepsc|s)^{\alpha} \Big)^{\frac{1}{\alpha}} - \nonumber \\
    &\qquad p(\Xepsc) \Big( \sum_{s} p(s) p(y'|s)^{\alpha} \Big)^{\frac{1}{\alpha}} \Big]\label{eq:MinkIneq2} > 0 %\label{eq:AssumptionIneq}
\end{align}
Equality \eqref{eq:MinkEq1} is obtained by substituting $r(y|x)$ by $1- \sum_{y' \in \Xepsc \colon y' \neq y}r(y'|x)$;
Inequality \eqref{eq:MinkIneq1} and \eqref{eq:MinkIneq2} are due to the Minkowski inequality.\footnote{Let $\| X \|_{\alpha} = (\E[|X|^\alpha])^{1/\alpha}$. By Minkowski's inequality,
$\|X\|_{\alpha} - \|Y\|_\alpha = \|X-Y+Y\|_{\alpha} - \|Y\|_\alpha \leq \|X - Y\|_\alpha$ and $\| \sum_{i = 1}^{n} X_i \|_{\alpha} \leq \sum_{i = 1}^{n} \|X_i\|_{\alpha}$. }
The last strict inequality is due to the assumption in the proof that for all $y' \in \Xepsc$ such that $y' \neq y$, we have $\ell_{\alpha}(y') < \bar{\ell}_{\alpha}$.

\bibliographystyle{IEEEtran}
\bibliography{BIB}

\end{document}